\documentclass[journal]{IEEEtran}

\usepackage{graphicx}
\usepackage{multirow}
\usepackage{amsmath,amssymb,amsfonts}
\usepackage{booktabs}
\usepackage{makecell}
\usepackage{textcomp}
\usepackage{xcolor}
\usepackage{enumitem}
\usepackage{url}
\usepackage{hyperref}
\usepackage[normalem]{ulem} 

\begin{document}

\title{Design and Evaluation of a Body-Machine Interface as an Alternative Control Paradigm for Pedal-based Sewing Machines}

\author{Ashrut~Aryal,
        Rajiv~Ranganathan,
        and~Ranjan~Mukherjee%
\thanks{This work was supported by NSF grant 2326227 and a Strategic Partnership Grant from Michigan State University. \textit{(Corresponding author: Rajiv Ranganathan.)}}%
\thanks{A.~Aryal, R.~Ranganathan, and R.~Mukherjee are with the Department of Mechanical Engineering, Michigan State University, East Lansing, MI 48824, USA (e-mail: aryalash@msu.edu; rrangana@msu.edu; mukherji@msu.edu).}}

\maketitle

\begin{abstract}
Body-machine interfaces (BoMIs) exploit body movements to control assistive devices; they have been evaluated primarily in tasks such as cursor control or robotic manipulation where the entire movement repertoire is devoted toward device control. However, tasks with multiple concurrent processes require the user to share the movement repertoire used for device control with other body movements for task execution. This issue is examined here in the context of sewing, where the sewing machine is typically controlled by a pedal in coordination with hand movements for guiding the fabric. We evaluated the learning of a trunk-controlled BoMI for sewing machine control, and compared it with conventional pedal control. Twenty-four novice participants (12 BoMI, 12 Pedal) completed 10 trials of a standardized sewing assessment followed by 3 trials of a transfer task. The BoMI mapped trunk movements to the sewing machine's full functional repertoire, which included continuous speed control of the needle, and discrete actions such as presser-foot and needle lifts. The two groups learned at comparable rates, showing similar completion times and error counts; for both groups, completion time reduced by approximately 30\%. Additional analyses showed that the BoMI group had fewer presser-foot lift errors during training and fewer errors on the first transfer trial. Both groups also exhibited similar subjective workload. These findings collectively demonstrate that a BoMI interface can support sewing-machine control during concurrent manual activity, highlighting the potential for using BoMIs for increasing accessibility in industrial workplace contexts.
\end{abstract}


\section{Introduction}
Body-machine interfaces (BoMIs) \cite{casadio2012body} translate voluntary body movements into commands for external devices, offering an alternative control mechanism when conventional inputs (such as joysticks or keypads) are limited or suboptimal. Developed primarily for assistive technology and neuromotor rehabilitation \cite{pierella2014,seanez2016}, BoMIs have been employed across a range of control complexities, from low-dimensional tasks such as cursor navigation \cite{seanez2013,abdollahi2017,seanez2016} to higher-dimensional applications including assistive robotic arm operation \cite{rizzoglio2023,lee2024, chau2017five,ranganathan2019age,aspelund2020controlling}, control of upper-limb prostheses and wearable exoskeletons \cite{scheme2011,fu2022,farina2014}, and drone teleoperation \cite{miehlbradt2018,yiu2025}. A consistent finding across these applications, in both movement-impaired and unimpaired populations, is that users can learn to reorganize their available movement repertoire to achieve proficient device control  \cite{pierella2014,seanez2016,lee2024}. In spite of this need to learn the interface,  BoMI-mediated control has often achieved comparable performance with conventional interfaces \cite{thorp2016,loboprat2014}, and in some applications has even exceeded it \cite{miehlbradt2018}.

However, prior BoMI studies have examined tasks in which the interface-controlled action constitutes the \textit{only} activity performed by the user (such as steering a wheelchair, or positioning a robotic end-effector). In these cases, the user's entire movement repertoire and attentional resources are devoted to learning the BoMI. In contrast, in many real-world industrial tasks, interface control rarely occurs in isolation; instead, interface control has to be coordinated with other concurrent motor actions. This coordination demand is compounded when the device requires hybrid control involving both continuous modulation (e.g.,  speed control), and discrete state transitions (e.g.,  on/off switching). This highlights the need for attentional and motor coordination between the part of the movement repertoire used for device control and the part used for other concurrent actions. In other words, users must divide a finite movement repertoire between interface control and the other concurrent actions while also managing the attentional demands of switching between continuous and discrete command modes. This ability to partition movement repertoire requires  the presence of redundancy so that movements for interface control do not affect the other task \cite{prattichizzo2021human, guggenheim2020leveraging}. Although prior studies have shown proof-of-concept that such control and coordination is possible \cite{kieliba2021robotic, parietti2017independent}, to our knowledge, no study has evaluated a BoMI in comparison to conventional interface in tasks requiring both interface-mediated hybrid control and a concurrent motor task.

Sewing serves as a prototypical example of a task requiring concurrent demands involving both machine control and fabric manipulation. The sewing machine consists of a needle that performs the stitching while a presser foot holds the fabric against the table. During sewing, the operator must continuously regulate needle speed (typically using a foot pedal) while using both hands to guide and tension the fabric along the desired path. Two additional operations must be performed using the foot pedal at zero needle speed: (a) lifting the presser foot to facilitate rotation of the fabric about the needle, and (b) lifting both the presser foot and the needle to facilitate positioning the fabric to a new location. Thus, sewing requires cycling between these three states and simultaneously performing bimanual manipulation specific to these states. Prior efforts to develop alternative control interfaces for sewing have addressed only portions of this hybrid control problem. Curry and Leamon \cite{curry1993} implemented a chest-operated lever that replaced the foot pedal for speed control on industrial machines, reporting equivalent accuracy and completion time in novice operators. However, that system provided only continuous speed modulation and did not implement the discrete presser foot and needle lift functions integral to standard sewing operations. More recent commercial solutions \cite{bernette_b37} offer start/stop buttons with preset speed dials, but these sacrifice the real-time proportional speed modulation needed to vary stitching rate during the task. Our goal was to develop a more versatile BoMI that allows control of the sewing machine in all three states, while allowing the user to simultaneously perform fabric manipulation.

In this study, we designed a BoMI that maps trunk movements to the full set of sewing machine commands: trunk flexion and return for proportional speed control, and trunk hyperextension (bending backwards beyond the neutral position) and return for cycling between the states associated with the presser foot lift and needle lift. We evaluated the BoMI by comparing it directly with the conventional foot pedal control. We used the Blum's test \cite{blum1943}, a standardized sewing skill assessment, to characterize learning trajectories. We compared the interfaces at multiple levels: task performance measured by time and errors, control behavior (that provided fine-grained level analyses of performance), and subjective workload. In addition, we used a transfer test to examine if the skills acquired during training generalized to a more complex task. This design is motivated in part by the need for alternative controls for users with limited lower-limb function, although the present study evaluated feasibility with unimpaired young adults. Our results show that no statistically detectable difference in overall training task performance was observed between BoMI and pedal-based control, indicating that users may effectively partition their movement repertoire during concurrent task execution.

\section{Materials and Methods}
\subsection{Participants}
Twenty-four college-aged adults (27 $\pm$ 5 years) with no prior sewing experience participated in this study.  Inclusion criteria required participants to be free from any musculoskeletal conditions affecting torso or lower limb mobility. Participants were paid \$15 for their participation in the study. All participants provided written informed consent, and the study protocol was approved by the Institutional Review Board at Michigan State University (IRB study number: STUDY00009492; approval date: 9/28/25). The sample size of $n=12$ per group is comparable to prior proof-of-concept BoMI studies \cite{rizzoglio2023, lee2024}. Additional allocation details and group-wise participant characteristics are provided in Supplementary Section~S1.

\subsection{Experimental Setup}
\begin{figure*}[t]
\centering
\includegraphics[width=\textwidth]{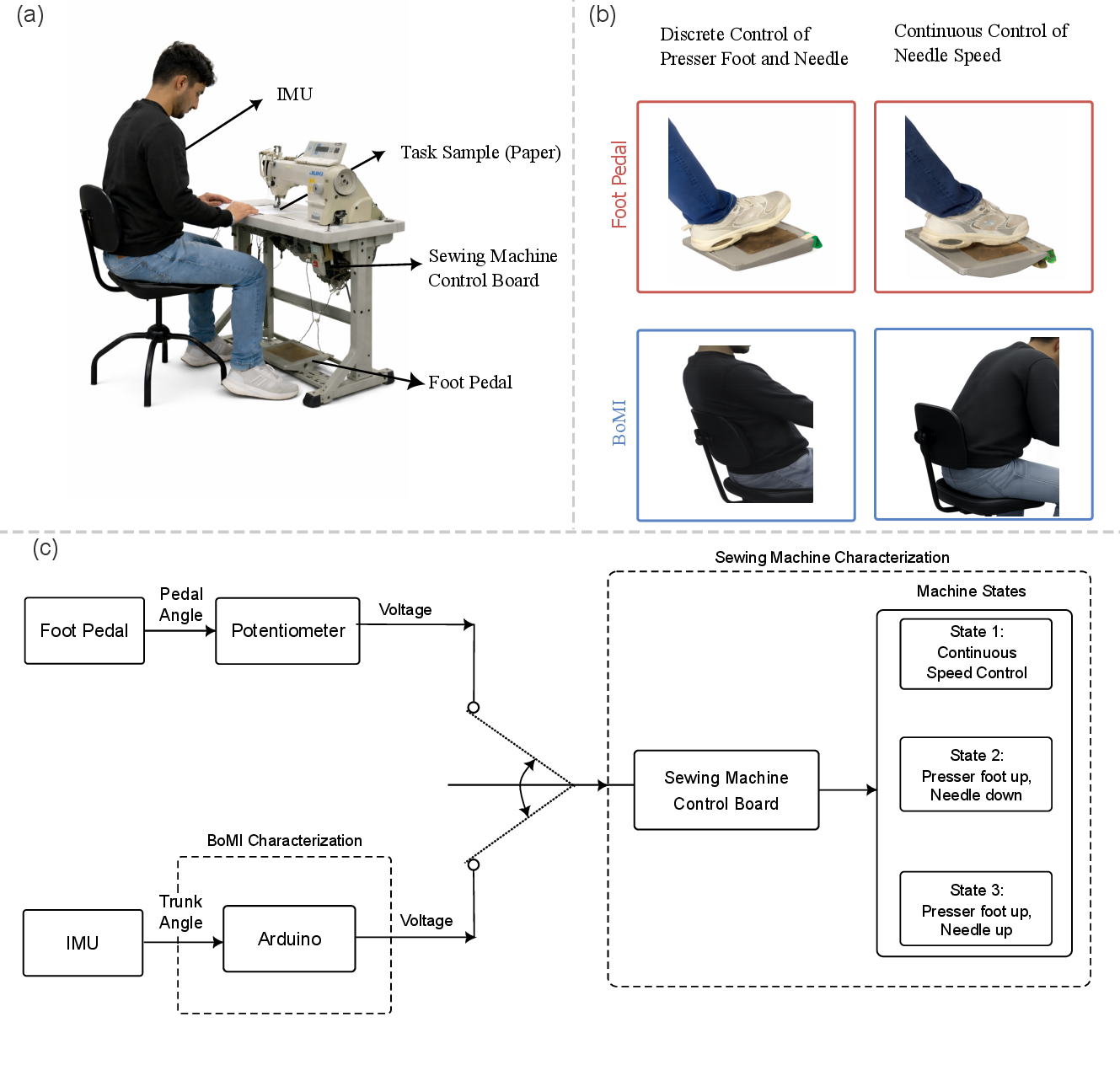}
\caption{Experimental setup for foot-pedal and BoMI control. (a) Experimental setup showing a participant seated at the sewing machine. (b) Sewing machine configuration for the two experimental groups---the machine could be controlled either using a foot pedal (top row) or through the BoMI by trunk movement (bottom row). (c) Sewing Machine Control architecture. Foot pedal input is converted to voltage via a potentiometer, while trunk angle is measured by an IMU and processed by an Arduino to generate control voltages. Both inputs are sent to the sewing machine control board to drive operation}
\label{fig1}
\end{figure*}

The experimental apparatus is shown in Fig.~\ref{fig1}(a). An industrial sewing machine (JUKI DDL-8700-7, Tokyo, Japan, \url{https://juki.com/ddl-8700-7-sc920c-m92}) was mounted on a standard worktable. The behavior of this machine is governed by a single analog voltage delivered to its control board, which governs functions such as the sewing machine speed, and discrete functions such as the presser foot lift and  needle lift. 

\subsection{Groups }
Participants were assigned into one of two groups - the foot pedal group ($n = 12$) or the BoMI group ($n = 12$) - see Fig.~\ref{fig1}(b). For the foot pedal group, the machine was operated using the built-in foot pedal. When the operator pressed the foot pedal (forward or backward), a potentiometer changes the voltage supplied to the control board of the machine - see Fig.~\ref{fig1}(c). 
For the BoMI group, the foot pedal was bypassed. Instead, a wireless Inertial Measurement Unit (IMU; Delsys Trigno IM sensor, Delsys Inc., USA) was affixed to the participant's upper sternum, and used to measure the participant's trunk angle. This angle was mapped to a command voltage using an Arduino Uno microcontroller, and sent to the control board - see Fig.~\ref{fig1}(c). Thus, the participant could lean forward and backward to control the sewing machine in a  similar manner to the foot pedal.

\begin{table}[h]
\centering
\caption{Sewing machine voltage and speed settings}
\label{tab:voltage_values}
\begin{tabular}{l|ccccccc}
\toprule
\text{Parameter} & $V_{NL}$ & $V_{PF}$ & $V_{\text{idle}}$ & $V_{1}$ & $V_{2}$ & $V_{3}$ & $V_{\text{max}}$ \\
\midrule
\text{Voltage (V)} & 1.15 & 1.6 & 1.9 & 2.24 & 2.66 & 3.78 & 4.4 \\
\midrule
\multicolumn{8}{l}{\footnotesize Measured minimum active sewing speed: 200 RPM} \\
\multicolumn{8}{l}{\footnotesize Measured maximum active sewing speed: 2054 RPM} \\
\bottomrule
\end{tabular}
\end{table}

\subsection{Sewing Machine Characterization}\label{subsec:machinechar}
\begin{figure*}
\centering
\includegraphics[width=\textwidth]{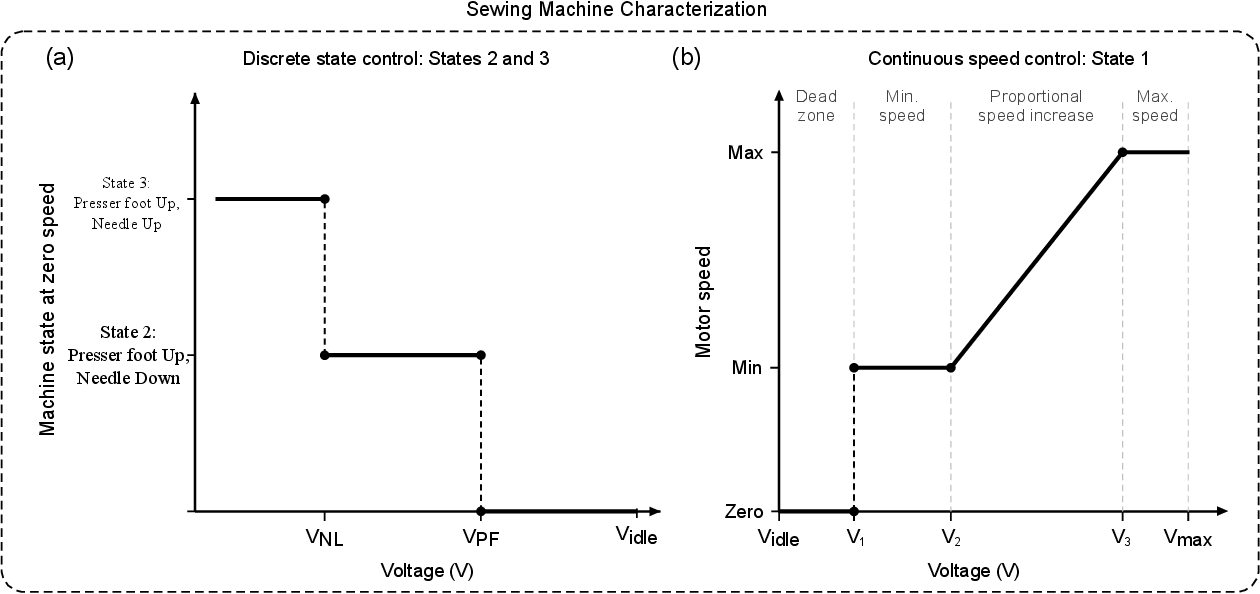}
\caption{Sewing machine characterization showing relationship between input voltage and machine behavior. (a) Discrete state control at zero speed showing transitions to the two discrete machine states—(PF up, needle up), (PF up, needle down)—as a function of voltage (below $V_{\mathrm{idle}}$). (b) Continuous speed control zone (above $V_{\mathrm{idle}}$) containing four operating regions: dead zone, minimum speed, proportional speed, and maximum speed.}
\label{fig2}
\end{figure*}
The behavior of the sewing machine is determined by voltage to the control board. We characterized this relationship by supplying a range of voltages to the control board via Arduino and recording the corresponding machine behavior. 

The machine's response to voltage is nonlinear and encompasses three states:

\begin{enumerate}[label=State \arabic*:, leftmargin=*, itemsep=0pt, topsep=2pt]
    \item For $V_{\text{idle}} \leq V \leq V_{\text{max}}$\footnote{Here, $V_{\text{idle}}$ is the voltage supplied to the control board when the pedal is in its neutral (not pressed) position, and $V_{\text{max}}$ is the maximum voltage supplied when the pedal is fully pressed.}, the machine is in the ``continuous speed control'' mode. As the voltage increases, the machine speed increases from zero to the maximum speed in four stages (see Fig.~\ref{fig2}(b)). While the zone with proportional speed increase requires precise movement control, the plateaus at the dead-zone, minimum speed, and maximum speed reduce the precision requirements. The plateau at the dead-zone extends to the ``discrete state control'' mode, and the speed remains zero for $V_{\text{PF}} \leq V \leq V_{\text{idle}}$ (see Fig.~\ref{fig2}(a)).
    
    \item For $V_{\text{NL}} \leq V \leq V_{\text{PF}}$, the machine speed is zero and the presser foot is lifted. In this state, the user can rotate the fabric about the needle.
    
    \item For $V \leq V_{\text{NL}}$, the machine speed is zero, the presser foot is lifted, and additionally the needle is lifted. In this state, the user can move the fabric freely.
\end{enumerate}

The above characterization shows that sewing machine control is a hybrid control problem with a single degree of freedom, wherein the operator must cycle between State 1: continuous speed control mode and States 2 and 3: discrete state control mode, for effective sewing. Furthermore, within each of these states, the operator must manipulate the fabric concurrently.

For the specific sewing machine used in this study, the different voltage and speed values are given in Table~\ref{tab:voltage_values}.

\subsection{BoMI Characterization}
The behavior of the BoMI is characterized by how the trunk angle is mapped into a voltage that is fed to the sewing machine control board. This design required three steps -- a calibration phase to determine the range of angles for a particular user, the design of the angle-to-voltage mapping in the continuous speed control state, and the design of the angle-to-voltage mapping in the discrete states. 

\subsubsection{Calibration}
Participants in the BoMI group initially completed a brief calibration procedure to establish their individual range of motion. Participants were asked to lean forward, as far as they could, comfortably, while being able to look at and manipulate the fabric. This angle measured by the IMU is the maximum trunk flexion $\theta_{\max}$. The participants then returned to a comfortable upright posture. This is the neutral trunk angle, $\theta_0$.
\begin{figure*}
\centering
\includegraphics[width=\textwidth]{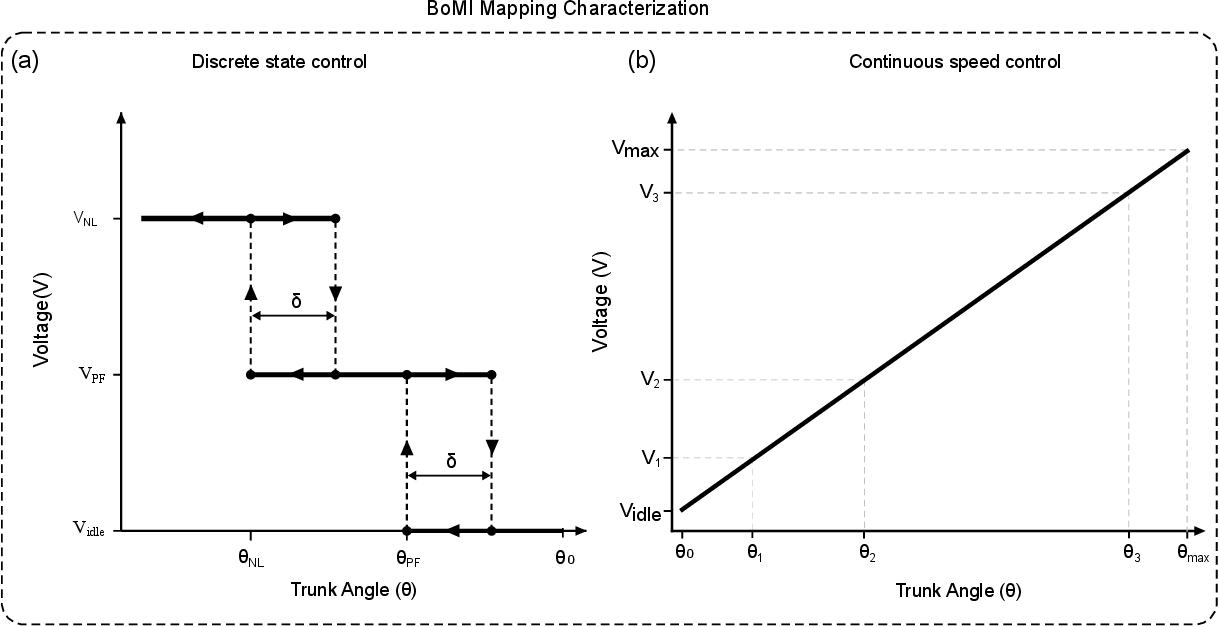}
\caption{BoMI mapping characterization showing relationship between trunk angle and voltage sent to the sewing machine. (a) Discrete state control: backward trunk lean beyond thresholds $\theta_{NL}$ and $\theta_{PF}$ commands voltages $V_{NL}$ and $V_{PF}$ respectively, triggering the discrete machine states (States 2 and 3), with hysteresis band $\delta$ preventing unintended state switching. (b) Continuous speed control: forward trunk flexion from the neutral angle $\theta_0$ to the maximum angle $\theta_{max}$ maps linearly onto the voltage range from $V_{idle}$ to $V_{max}$, with fixed breakpoints $\theta_1$, $\theta_2$, and $\theta_3$ corresponding to the machine's intrinsic speed zones.}
\label{fig3}
\end{figure*}

\begin{table*}[ht]
\centering
\caption{Angle Parameters for BoMI Characterization}
\label{tab:bomi characterization}
\begin{tabular}{l|ccc|ccccc}
\toprule
 & \multicolumn{3}{c|}{\text{User Independent}} & \multicolumn{5}{c}{\text{User Specific}} \\
\midrule
\text{Angle} & $\theta_{NL}$ & $\theta_{PF}$ & $\delta$ & $\theta_0$ & $\theta_1$ & $\theta_2$ & $\theta_3$ & $\theta_{\max}$ \\
\midrule
\text{Value} & $-8^\circ$ & $-4^\circ$ & $2^\circ$ & \makecell{Measured} & $\theta_0$ + $0.14\Delta\theta$ & $\theta_0$ + $0.30\Delta\theta$ & $\theta_0$ + $0.75\Delta\theta$ & \makecell{Measured} \\
\bottomrule
\multicolumn{9}{l}{\footnotesize $\Delta\theta = \theta_{\max} - \theta_0$}
\end{tabular}
\end{table*}

\subsubsection{Mapping in Continuous Speed Control State}
The trunk flexion (i.e., forward lean relative to the neutral posture $\theta_0$) was used to trigger the continuous speed control mode in a user-dependent manner. Based on the initial calibration procedure, the participant's usable trunk flexion range of motion was defined as:
\begin{equation}
\Delta \theta = \theta_{\max} - \theta_0
\end{equation}

\noindent The instantaneous trunk flexion angle $\theta$(t) was mapped to sewing machine command voltage, $V(t)$, as:
\begin{equation}
V(t) = V_{\mathrm{idle}} + \frac{\theta(t) - \theta_0}{\Delta \theta}(V_{\max} - V_{\mathrm{idle}})
\end{equation}
where $V_{\mathrm{idle}}$ is the idle voltage, and $V_{\mathrm{max}}$ is the maximum voltage. Because this mapping spans the entire voltage range linearly - see Fig.~\ref{fig3}(b), the nonlinear operating regions characterized in Section~\ref{subsec:machinechar} are inherited directly from the machine's own voltage-speed response.

For a specific machine, $V_1$, $V_2$, and $V_3$ are known apriori. Therefore the trunk angles $\theta_1$, $\theta_2$, and $\theta_3$ corresponding to these voltages are fixed for a user, given $\theta_0$ and $\theta_{\text{max}}$. 

\subsubsection{Mapping in Discrete States}
The trunk hyperextension (i.e., backward lean relative to the neutral posture $\theta_0$) was used to trigger the sewing machine's discrete states (States 2 and 3) in a user-independent manner. For all $\theta(t) < \theta_0$, the commanded voltage could assume only one of three discrete values:
\begin{equation}
V(t)\in\left\{V_{\mathrm{idle}},\,V_{\mathrm{PF}},\,V_{\mathrm{NL}}\right\},
\end{equation}
where $V_{\mathrm{PF}}$ denotes the voltage that triggers presser-foot lift, and $V_{\mathrm{NL}}$ denotes the voltage that triggers simultaneous presser-foot and needle lift. Thus, regardless of the actual angle, any backward lean relative to neutral corresponded only to one of three states: an idle state, the presser-foot-lift state (State 2), or the combined presser-foot-and-needle-lift state (State 3) - see Table \ref{tab:bomi characterization}.

To reduce chatter, i.e., unintended switching between discrete states caused by sensor noise or minor postural fluctuations, transitions between these discrete states were implemented using hysteresis - see Fig.~\ref{fig3}(a). Let $\theta_{\mathrm{PF}}$ denote the entry threshold for the presser-foot-lift state, and let $\theta_{\mathrm{NL}}$  denote the entry threshold for the combined presser-foot-and-needle-lift state. When $\theta(t)$ decreased below $\theta_{\mathrm{PF}}$, the controller output switched to $V_{\mathrm{PF}}$ and remained there until $\theta(t)$ increased above $\theta_{\mathrm{PF}} + \delta$. Likewise, when $\theta(t)$ decreased below $\theta_{\mathrm{NL}}$, the controller output switched to $V_{\mathrm{NL}}$ and remained there until $\theta(t)$ increased above $\theta_{\mathrm{NL}} + \delta$. If neither threshold condition was satisfied, the controller was set to $V_{\mathrm{idle}}$.

Additional details of IMU acquisition, signal processing, command transmission, and controller safety are provided in Supplementary Section~S2.

\subsection{Task Protocol}
All participants, regardless of group assignment, completed an identical sequence of three tasks: a familiarization task with 3 trials, a training task with 10 trials, and a transfer task with 3 trials. All sewing tasks in this experiment were performed on a standard letter-sized (8.5 x 11 inch) paper with no thread. This enabled precise quantification of errors by examining the needle perforations on the paper at the end of each trial. Other than the control interface (including its controller logic which includes hysteresis for the BoMI), all other sewing machine settings were kept the same for all participants.

\begin{figure*}
\centering
\includegraphics[width=\textwidth]{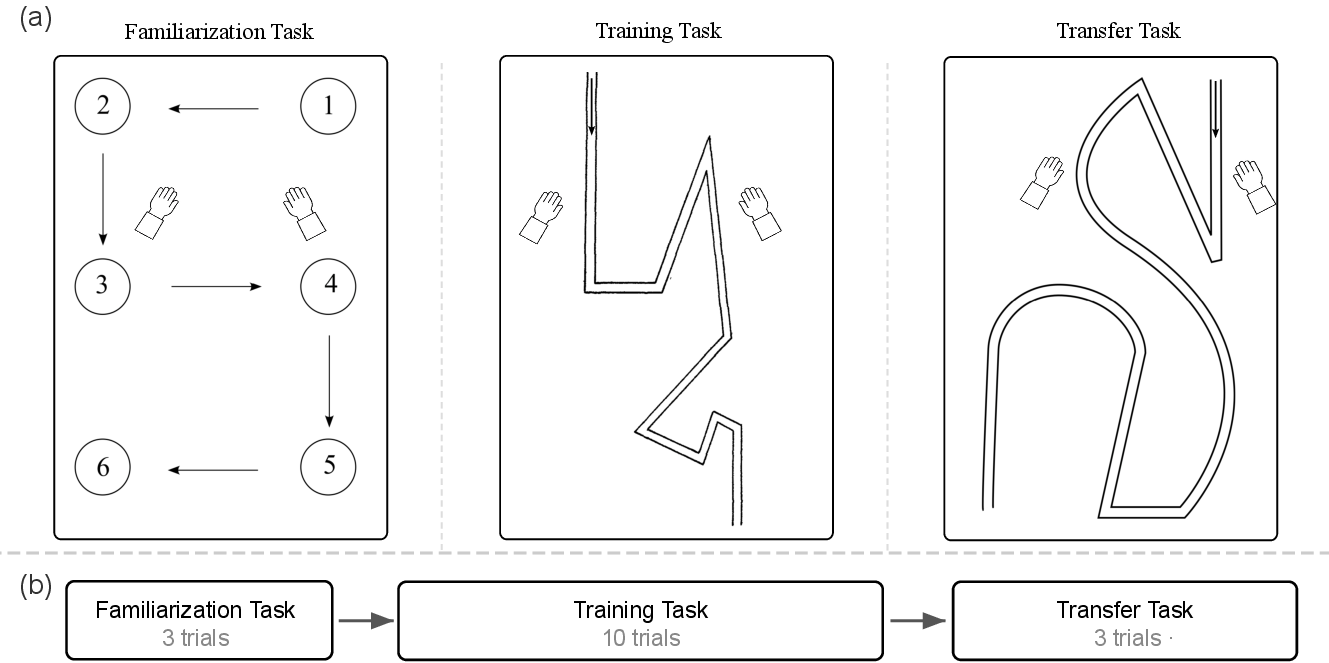}
\caption{(a) Task templates (printed on letter-sized paper) used for familiarization, training, and transfer tasks. Arrows indicate sewing direction. Hands are plotted (not to scale) to indicate the orientation of the paper relative to the participant at the beginning of each task.
(b) Experimental protocol. All participants completed three phases in sequence: familiarization (3 trials), training (10 trials), and transfer (3 trials).}
\label{fig4}
\end{figure*}

\subsubsection{Familiarization}
Since participants did not have prior sewing experience,  the familiarization phase was designed to make sure that participants were familiar with the control interface (foot pedal or BoMI) and basic sewing operations. This phase began with a standardized orientation to the sewing machine. The experimenter explained and demonstrated the basic functions: starting and stopping the needle, controlling speed, lifting the presser foot to reorient the paper, and lifting the needle to remove the paper at the end of the trial.

Participants then completed three trials of the familiarization task - see Fig.~\ref{fig4}(a), left panel. In this task, participants placed the paper under the needle with the needle positioned anywhere within the starting circle (labeled ``1''), then sewed to the next circle (labeled ``2''), aiming to stop with the needle inside that circle. At each circle, participants lifted the presser foot only (i.e., State 2), rotated the fabric to align the sewing direction to the next circle, lowered the presser foot (transitioned back to State 1), and resumed sewing. This sequence started from circle 1 and ended at circle 6. At the end of each trial, both presser foot and needle, were lifted (i.e., State 3) to remove the paper.

\subsubsection{Training}
The training task was a standardized sewing skill assessment known as the Blum's test \cite{blum1943}, which requires stitching through a constrained zig-zag path -see Fig.~\ref{fig4}(a), center panel. The test template consisted of two parallel zig-zag lines forming a channel approximately 1.5 cm wide, with eight directional changes requiring paper rotations (which required cycling between State 1 and State 2). For example, the first rotation required a 90$^\circ$ clockwise turn about the vertical axis. Participants began at the designated start arrow (bottom of the template) and sewed along the passage, attempting to keep all needle perforations within the two parallel lines. At each corner where the path direction changed, participants were required to stop sewing, lift the presser foot (transition to State 2), rotate the fabric to align with the next path segment, lower the presser foot (transition back to State 1), and resume sewing. Participants were instructed to go as fast as possible while minimizing error (number of needle perforations outside the channel, described in Section~\ref{subsec:taskperformance}).

Each participant completed 10 consecutive trials. After each trial, the experimenter recorded the completion time and error, and provided the participants with feedback on both measures.

\subsubsection{Transfer}
For the transfer task, participants performed three trials of a modified version of the Blum's test. The rationale for this modification was to assess whether learning from the strictly stop-start Blum's test transferred to more complex, dynamic fabric manipulation scenarios common in actual garment sewing - see Fig.~\ref{fig4}(a), right panel. The modified template consisted of four distinct segments: (1) a straight entry segment, (2) an angled segment similar to the standard Blum's test requiring a paper rotation of 150$^\circ$ counterclockwise, (3) a S-shaped segment requiring simultaneous speed and fabric angle adjustments without stopping, and (4) a U-shaped segment followed by a straight exit.
Each participant completed 3 consecutive trials. Evaluation and feedback procedures were identical to the standard Blum's test.

\section{Data Analysis}
\subsection{Evaluation Metrics}
Performance was assessed at three levels: task performance, control behavior, and subjective workload. Task performance metrics were computed for both the training and transfer tasks. Control behavior metrics, which were used to understand how performance was achieved, were computed for the training task only. Subjective workload was assessed following completion of the entire experiment.

\subsubsection{Task Performance}\label{subsec:taskperformance}
Given that the goal of the participant was to complete the task as quickly as possible without making errors, performance was quantified in terms of error count and completion time.

\textbf{Error Count.} Error count was defined as the number of needle perforations falling entirely outside the designated boundary lines of the template and was determined through visual inspection \cite{blum1943}. Lower errors indicate better task performance. A blinded second rater independently rescored a randomly selected subset of task templates; details of reliability assessment are provided in Supplementary Section~S3.

\textbf{Completion Time.} Completion time was measured using a stopwatch from the moment the participant began the first stitch until completing the final stitch at the end of the template, recorded to the nearest second. Lower completion times indicate better task performance.  Additional timing-validation details are provided in Supplementary Section~S3.

\subsubsection{Measures of Control Behavior}
Beyond overall task performance, we sought to understand the mechanisms through which performance was achieved. To this end, we examined the underlying control behavior during training using three measures.

\textbf{Mode Errors.} To characterize the participant's ability to reliably navigate the discrete control space, we examined the number of mode errors during training. The Blum's test template requires eight presser foot lifts and zero needle lifts. Therefore, we measured: (i) presser foot lift errors, and (ii) needle lift errors as the number of excess presser-foot lifts beyond the eight required lifts and the number of unintended needle lifts, respectively. Lower errors indicate better performance. No required presser-foot lifts were omitted in the available valid training trials.

\textbf{Time distribution.} Using the machine voltage characterization described in Section~\ref{subsec:machinechar} (Fig.~\ref{fig2}), we computed the proportion of total trial time spent in each of six mutually exclusive operating regions: dead zone, minimum speed zone, proportional speed zone, maximum speed zone, presser foot lift zone, and needle and presser foot lift zone. We also computed the sewing time as the combined proportion of time spent in the proportional speed, minimum speed, and maximum speed zones---i.e., the time in which the needle was moving.

\textbf{Mean Sewing Speed.} To characterize how sewing speed evolved with practice, we computed the mean  speed (RPM) across the sewing states (when the needle was moving).  Active sewing speed was obtained from the recorded voltages using the characterized voltage--speed relationship.

\subsubsection{Subjective Workload}
Following completion of all experimental trials, participants completed the NASA Task Load Index (NASA-TLX) \cite{hart1988}, a widely used instrument for assessing perceived workload across six dimensions: mental demand, physical demand, temporal demand, performance, effort, and frustration. This measure was used to compare the usability of the two interfaces \cite{gentili2015cognitive}. Lower values indicate lesser workload (i.e., better rating of user interface).

\subsection{Statistical Analysis}

Training completion time was designated as the primary outcome. Transfer performance (completion time and errors), and subjective workload (NASA-TLX) were classified as secondary outcomes, as these were part of the planned evaluation of transfer performance and subjective workload. Training errors, presser-foot lift errors, needle-lift errors, percentage of sewing time, and mean active sewing speed were classified as exploratory outcomes, as these measures provided additional information to characterize task accuracy and control behavior.

\noindent \textbf{Training.}
For training, we used mixed-effects models with Group (BoMI, Pedal), Trial, and their interaction as fixed effects and participant as a random intercept. Trial was treated as a categorical variable. Completion time was analyzed using a linear mixed-effects model (LMM) following a natural-log transformation to reduce residual skewness. The three training count outcomes (errors, presser-foot lift errors, and needle-lift errors) were analyzed using negative-binomial generalized linear mixed-effects models (GLMMs) with a log link to accommodate overdispersion, and model effects were evaluated using likelihood-ratio tests. The supporting control-behavior measures (percentage of sewing time and mean active sewing speed) were similarly analyzed using LMMs, with a natural-log transformation applied to mean active sewing speed to reduce residual skewness.

To account for multiplicity among exploratory training outcomes, Holm's correction was applied across 15 omnibus tests corresponding to the Group, Trial, and Group $\times$ Trial effects for each of the five exploratory outcomes (errors, presser-foot lift errors, needle-lift errors, percentage of sewing time, and mean active sewing speed).

\noindent \textbf{Transfer.}
For transfer, Trial~1 was used as the principal measure of immediate transfer. Between-group differences in completion time and error counts were analyzed using a linear model on log-transformed time and a negative-binomial generalized linear model (GLM), respectively. Performance across all three transfer trials was additionally analyzed using a log-time LMM and a negative-binomial GLMM, with Group, Trial, and their interaction as fixed effects and participant as a random intercept. Holm's correction was applied across eight comparisons: the two between-group comparisons on Transfer Trial~1 (completion time and errors), and the Group, Trial, and Group $\times$ Trial effects from the longitudinal models for both outcomes.

\noindent \textbf{Subjective workload.} For each of the six NASA-TLX subscales (mental demand, physical demand, temporal demand, performance, effort, and frustration), between-group differences were tested using Mann--Whitney $U$ tests. Holm's correction was applied across the six subscale comparisons. For each subscale, the BoMI--Pedal mean difference is reported with a Welch 95\% confidence interval.

\noindent For all analyses, the level of significance was set at $\alpha = 0.05$. Statistical analyses were performed in JASP \cite{love2019jasp} and Python, using pandas \cite{mckinney2010pandas} and NumPy \cite{harris2020array} for data processing, and SciPy \cite{virtanen2020scipy} and statsmodels \cite{seabold2010statsmodels}, for statistical testing.

Model results are reported with effect estimates, 95\% confidence intervals, and $p$ values. Exponentiated group contrasts from log-transformed models are expressed as ratios, and those from count models as rate ratios.

Equivalence and sample-size sensitivity analyses and detailed transfer analyses are provided in Supplementary Sections~S4 and~S5, respectively. Unadjusted and Holm-adjusted $p$ values for exploratory training, transfer, and NASA-TLX analyses are reported in Supplementary Section~S7.

\section{Results}

The voltage log was unusable for one participant in the Pedal group. Therefore, the analyses involving measures of control behavior had data from only 23 participants.

\subsection{Training}

\subsubsection{Completion Time}
Both groups exhibited substantial improvement over training (Fig.~\ref{fig5}a). The LMM on completion time showed a significant Trial effect, $F(9,198)=45.27$, $p<.001$, confirming that participants in both groups became markedly faster with practice. There was no statistically detectable Group effect, $F(1,22)=0.019$, $p=.892$, or Group $\times$ Trial interaction, $F(9,198)=1.00$, $p=.445$. The estimated BoMI-to-Pedal completion-time ratio was 0.987 (95\% CI: 0.814--1.196). In terms of magnitude of improvements, the completion time from the first three to the last three trials showed reductions of 33\% and 32\% for the BoMI and the Pedal groups respectively. 

\subsubsection{Error Count}
\begin{figure*}[t]
\centering
\includegraphics[width=\textwidth]{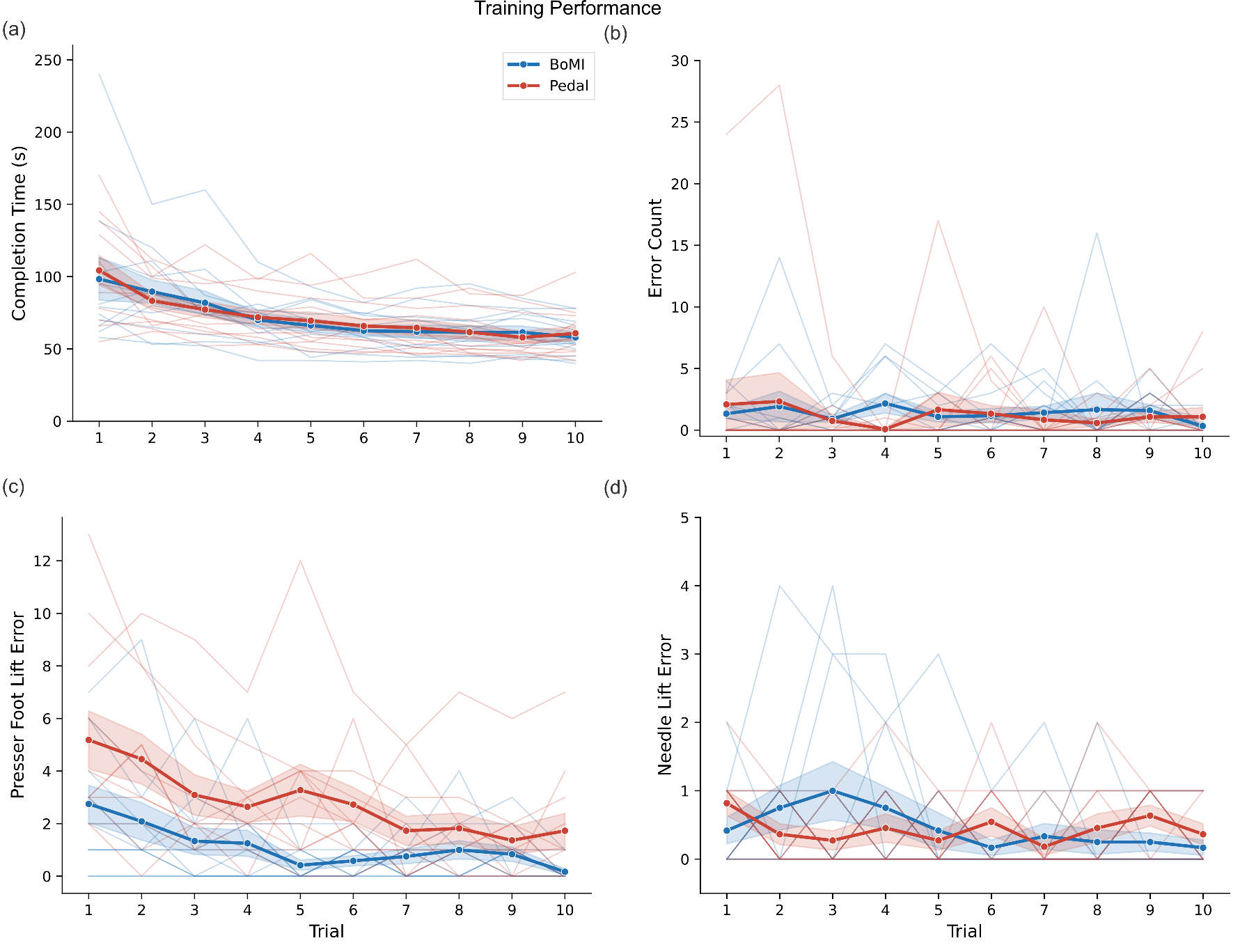}
\caption{Training performance across the 10 Blum's test trials for the BoMI (blue) and Pedal (red) groups. Thick lines show group means shaded $\pm$1 SEM bands; thin lines show individual participant trajectories.  (a) Completion time. (b) Error count. (c) Presser foot lift error. (d) Needle lift error.}
\label{fig5}
\end{figure*}

Error rates were low throughout training in both groups (Fig.~\ref{fig5}b). The negative-binomial GLMM found no statistically detectable Group effect, $\chi^2(1)=2.00$, Holm-adjusted $p=1.000$; Trial effect, $\chi^2(9)=10.71$, Holm-adjusted $p=1.000$; or Group $\times$ Trial interaction, $\chi^2(9)=8.83$, Holm-adjusted $p=1.000$. The estimated BoMI-to-Pedal rate ratio was 2.319 (95\% CI: 0.727--7.391). Taken together with the completion time results, these findings indicate that learning in the BoMI was similar to the conventional foot pedal on the task-level performance measures.

\subsubsection{Measures of Control}

\textbf{Mode errors.} Both groups decreased presser foot lift errors with practice (Fig.~\ref{fig5}c), with the BoMI group showing a distinct advantage. The negative-binomial GLMM showed a Trial effect, $\chi^2(9)=96.58$, Holm-adjusted $p<.001$, and a Group effect, $\chi^2(1)=10.36$, Holm-adjusted $p=.0154$, with fewer errors in the BoMI group (BoMI-to-Pedal rate ratio 0.322, 95\% CI: 0.171--0.606). The Group $\times$ Trial interaction, $\chi^2(9)=19.59$, was not statistically significant (Holm-adjusted $p=.1853$).

    Needle lift errors were low throughout training in both groups (Fig.~\ref{fig5}d). The negative-binomial GLMM found no statistically detectable Group effect, $\chi^2(1)=0.39$, Holm-adjusted $p=1.000$; Trial effect, $\chi^2(9)=8.55$, Holm-adjusted $p=1.000$; or Group $\times$ Trial interaction, $\chi^2(9)=15.83$, Holm-adjusted $p=.5643$. The estimated BoMI-to-Pedal rate ratio was 0.770 (95\% CI: 0.338--1.755).

\begin{figure*}[t]
\centering
\includegraphics[width=\textwidth]{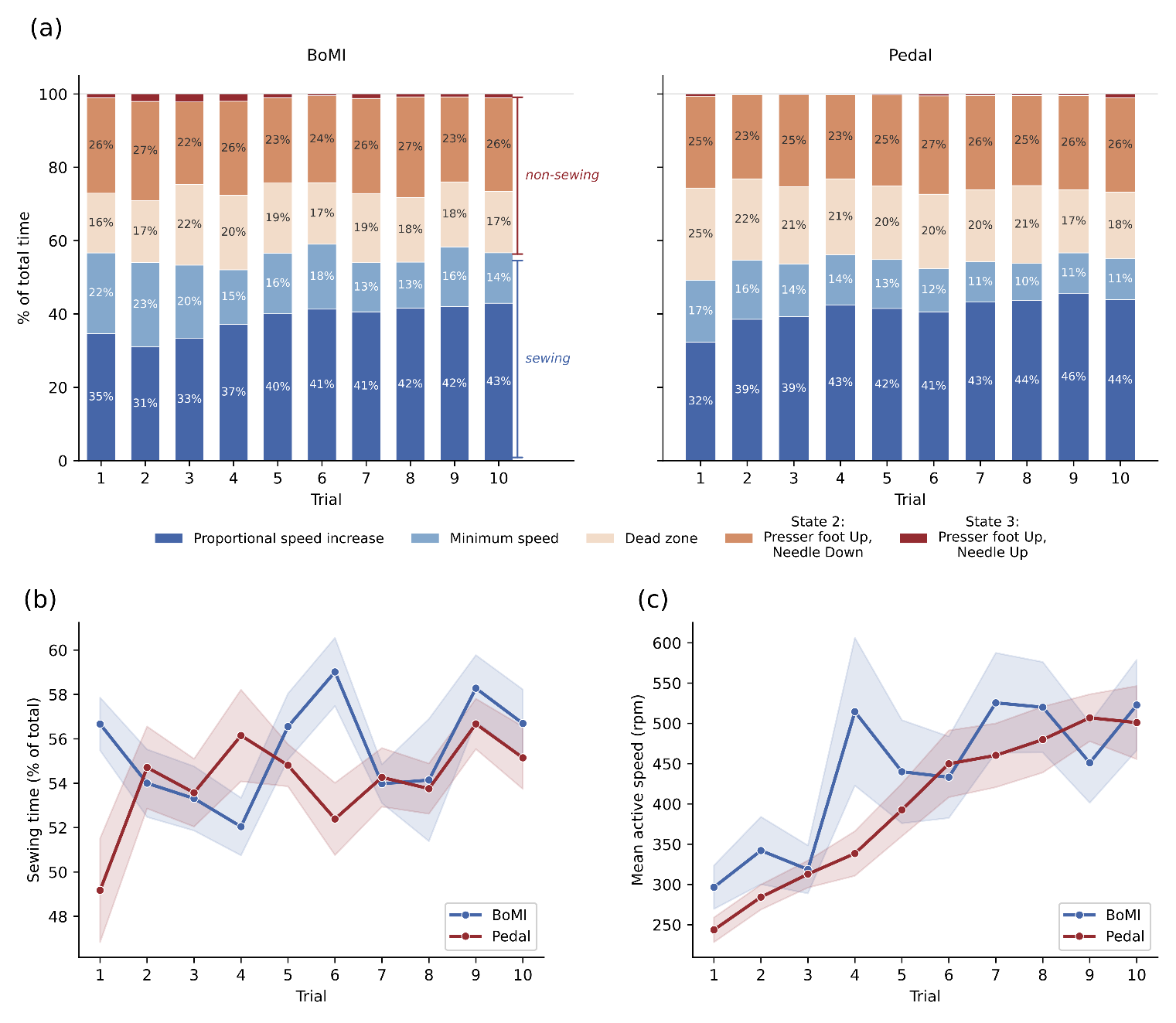}
\caption{Characterization of control behavior during training. (a) Time distribution as a percentage of total trial duration, shown separately for the BoMI (left) and Pedal (right) groups. Each bar illustrates the time spent in sewing and non-sewing states. Sewing states consisted of the proportional speed and minimum speed, and the non-sewing states consisted of the dead zone, presser foot lift, and needle \& presser foot lift. Percentages are group means per trial. (b) Mean percentage of total trial time spent sewing (proportional speed region + minimum speed region) across trials for both groups. (c) Mean active sewing speed (RPM) across trials. Shaded bands indicate $\pm$1 SEM. Both groups increased their mean sewing speed with practice.}
\label{fig6}
\end{figure*}

\textbf{Time distribution.} The two groups showed similar patterns of time allocation across machine states (Fig.~\ref{fig6}a). The LMM on percentage of sewing time (Fig.~\ref{fig6}b) showed no statistically detectable Group effect, $F(1,21)=0.89$, Holm-adjusted $p=1.000$ (estimated BoMI--Pedal difference 1.42 percentage points, 95\% CI: $[-1.53,\,4.37]$). The Trial effect, $F(9,189)=2.48$, was not statistically significant (Holm-adjusted $p=.1056$), whereas the Group $\times$ Trial interaction, $F(9,189)=3.80$, was significant (Holm-adjusted $p=.0026$). This interaction reflected trial-to-trial differences without a consistent direction; the largest difference occurred on Trial~1 (BoMI 56.7\%, Pedal 49.2\%), and the groups differed by less than 2 percentage points over the last three trials. It is worth noting that neither of the two groups operated at the maximum speed zone at any point during training, indicating that this result was not a ceiling effect because of reaching the machine's maximum speed.

\textbf{Mean sewing speed.} Both groups increased sewing speed with practice. The LMM on mean sewing speed showed a significant Trial effect, $F(9,189)=30.13$, Holm-adjusted $p<.001$. The Group $\times$ Trial interaction, $F(9,189)=2.60$, was not statistically significant (Holm-adjusted $p=.0839$). There was no statistically detectable Group effect, $F(1,21)=0.24$, Holm-adjusted $p=1.000$ (BoMI-to-Pedal speed ratio 1.060, 95\% CI: 0.840--1.338; Fig.~\ref{fig6}c).
Additional descriptive characterization of trunk control in the BoMI group is included in Supplementary Section~S6.




\subsection{Transfer Performance}

\begin{figure}[t]
\centering
\includegraphics[width=0.8\columnwidth]{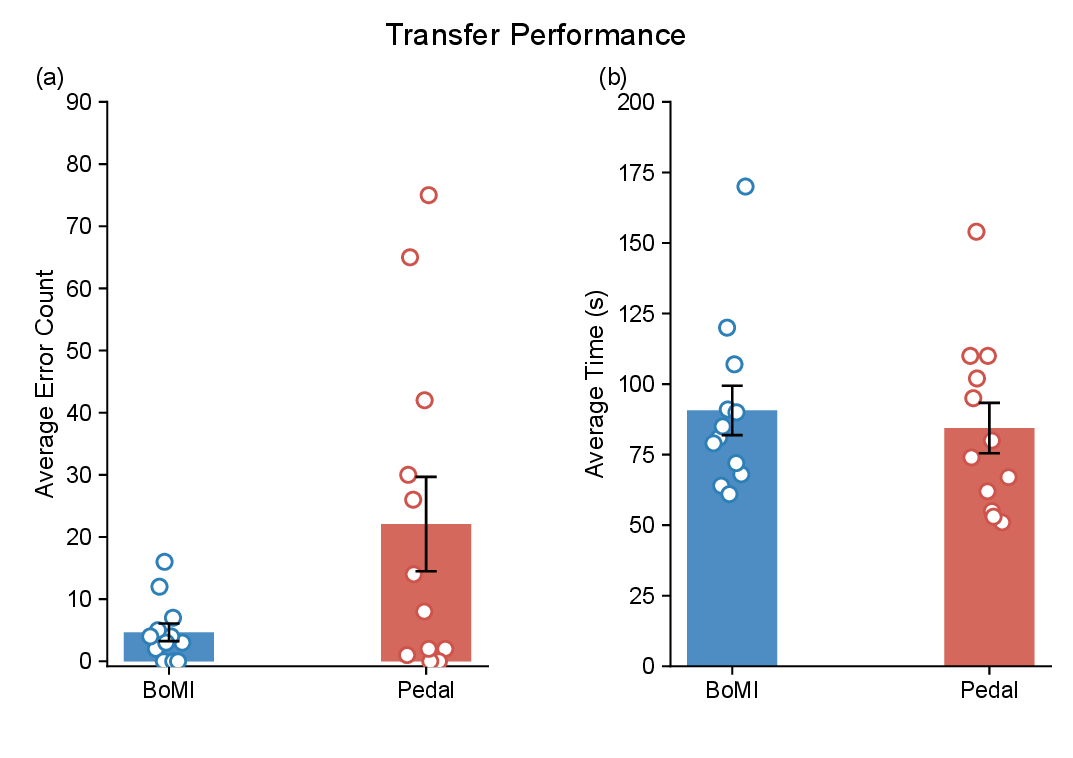}
\caption{Transfer performance. Group means with $\pm$1 SEM error bars; individual participant data points are overlaid. Transfer performance is shown for Trial~1. (a) Error count. The BoMI group produced fewer errors on the transfer task. (b) Completion time. Both groups showed comparable transfer times.}
\label{fig7}
\end{figure}

On the first transfer trial, BoMI participants made fewer errors (mean $4.67\pm4.92$) than Pedal participants (mean $22.08\pm26.30$) (Fig.~\ref{fig7}); the negative-binomial GLM gave a BoMI-to-Pedal rate ratio of 0.211 (95\% CI: 0.071--0.625, Holm-adjusted $p=.0368$). For completion time, no statistically detectable Group difference was observed (BoMI $90.67\pm30.36$~s; Pedal $84.42\pm30.86$~s; log-time model ratio 1.090, 95\% CI: 0.831--1.430, Holm-adjusted $p=1.000$).

Across all three transfer trials, mean errors in the Pedal group decreased from 22.08 on Trial~1 to 1.83 on Trial~3, whereas the BoMI group means were 4.67 and 3.33, respectively. For errors, the negative-binomial GLMM showed a significant Trial effect, $\chi^2(2)=10.47$, Holm-adjusted $p=.0368$, but no statistically detectable Group effect, $\chi^2(1)=2.68$, Holm-adjusted $p=.3042$, or Group $\times$ Trial interaction, $\chi^2(2)=5.16$, Holm-adjusted $p=.3037$. For completion time, the log-time LMM showed a significant Group $\times$ Trial interaction, $F(2,44)=6.10$, Holm-adjusted $p=.0368$: mean completion time in the BoMI group decreased from 90.67~s on Trial~1 to 75.83~s on Trial~3, whereas in the Pedal group it increased from 84.42~s to 95.75~s on Trial~2 before decreasing to 88.33~s on Trial~3. There was no statistically detectable Group effect, $F(1,22)=0.16$, Holm-adjusted $p=1.000$, or Trial effect, $F(2,44)=3.00$, Holm-adjusted $p=.3010$. Detailed longitudinal results are provided in Supplementary Section~S5, and unadjusted and Holm-adjusted $p$ values are reported in Supplementary Section~S7.2.

\subsection{Subjective Workload (NASA-TLX)}

\begin{figure}[t]
\centering
\includegraphics[width=\columnwidth]{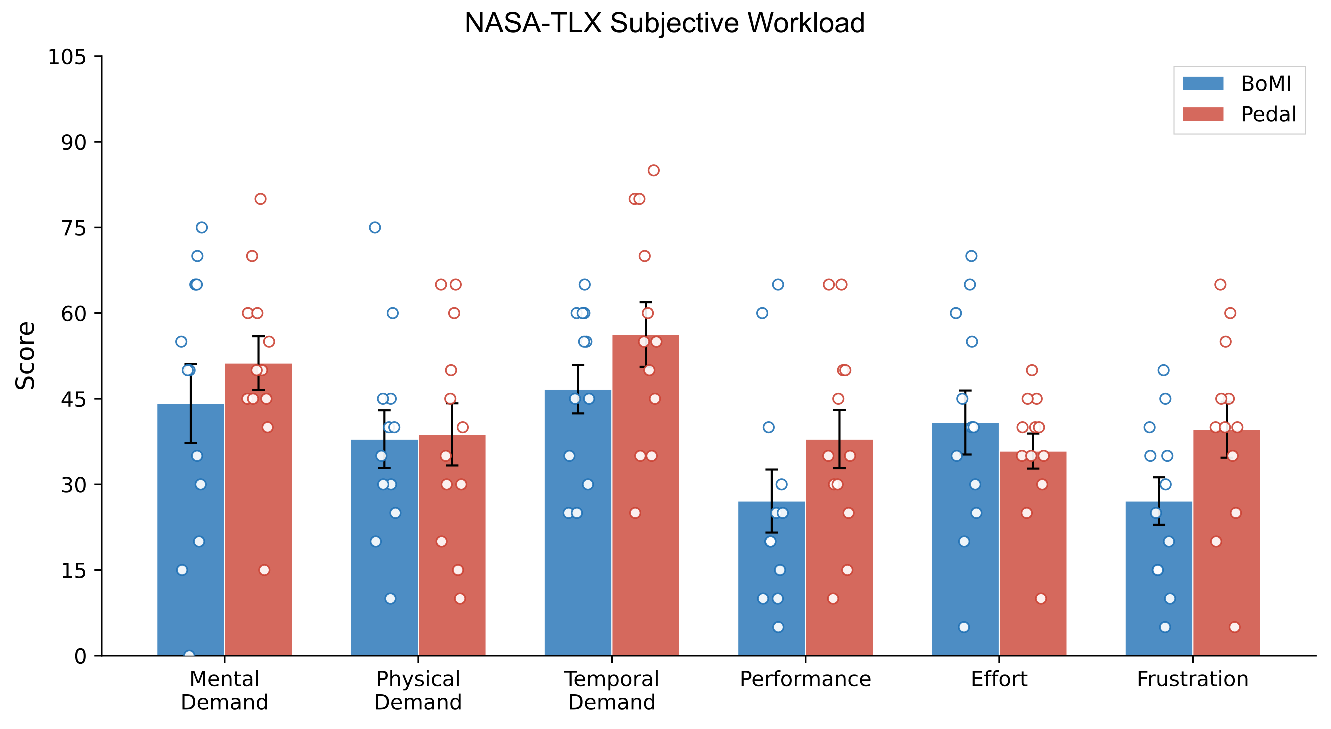}
\caption{NASA-TLX subjective workload ratings shown for the BoMI (blue) and Pedal (red) groups. Bars indicate group means with $\pm$1 SEM error bars; individual participant ratings are overlaid as open circles. No subscale showed a significant between-group difference. Lower scores indicate lower perceived workload.}
\label{fig8}
\end{figure}

NASA-TLX ratings are presented in Figure~\ref{fig8}. Both groups reported moderate overall workload, and no subscale showed a statistically significant between-group difference (mental demand: $U = 64.5$, Holm-adjusted $p = 1.000$; physical demand: $U = 69.0$, Holm-adjusted $p = 1.000$; temporal demand: $U = 53.5$, Holm-adjusted $p = 1.000$; performance: $U = 43.5$, Holm-adjusted $p = .522$; effort: $U = 82.5$, Holm-adjusted $p = 1.000$; frustration: $U = 40.0$, Holm-adjusted $p = .406$).
The estimated mean differences (BoMI minus Pedal) and their 95\% CIs were: mental demand, $-7.08$ [$-24.58$, $10.42$]; physical demand, $-0.83$ [$-16.24$, $14.57$]; temporal demand, $-9.58$ [$-24.28$, $5.12$]; performance, $-10.83$ [$-26.41$, $4.75$]; effort, $5.00$ [$-8.47$, $18.47$]; and frustration, $-12.50$ [$-25.84$, $0.84$].
This suggests that there was no statistically detectable difference in perceived workload between the interfaces.

\section{Discussion}
The purpose of this study was to determine whether a BoMI could support learning and performance comparable to a conventional foot pedal during sewing-machine control, a task requiring hybrid continuous-discrete machine operation together with concurrent bimanual fabric manipulation. Three principal findings emerged: (1) both groups improved in completion time during training, with no statistically detectable between-group differences in overall completion time or error count, (2) there was an advantage for the BoMI group in terms of fewer presser-foot lift errors and fewer errors in the immediate transfer task, and (3) there was no statistically detectable difference in subjective workload between the BoMI and the pedal interfaces. It is important to note that although the groups were similar in many variables, the study was not powered to establish equivalence.

\subsection{BoMI control achieved comparable performance to pedal control }
In spite of the relative novelty of the BoMI interface, both interfaces showed similar performance on the training task. The similarity in completion time and error count suggests that novice users were able to learn trunk-based machine control while simultaneously performing manual paper manipulation. The similarity in performance was seen at multiple levels in terms of the time spent in different actions, and the mean sewing speed.

The similarity between the interfaces is particularly noteworthy given the substantial differences in terms of the biomechanics and the sensory feedback between the two interfaces. The foot has lower inertia compared to the trunk, and participants in the pedal group receive sensory feedback from contact with the pedal interface, whereas the BoMI group performs these motions without contact. Despite these differences, users learned to control the sewing machine during concurrent manual activity. Moreover, there was no statistically detectable difference in perceived workload between the interfaces. This result is consistent with prior human-machine interface studies showing that, when task demands are held constant, different control modalities can yield comparable task-level performance. For example, Lobo-Prat et al. \cite{loboprat2014} reported equivalent performance between joystick-based and force-based interfaces in a standardized tracking task across multiple performance measures.

\subsection{BoMI showed fewer presser-foot lift and immediate-transfer errors}
Two outcomes favored the BoMI group: (a) presser-foot lift errors during training, and (b) errors on the first transfer trial.

The advantage in terms of the presser foot lift errors, which require triggering of discrete action states, may be partly attributable to the hysteresis incorporated into the BoMI design. We introduced a small tolerance around the presser-foot-up state, which meant that minor unintentional trunk movements near the threshold did not immediately trigger a discrete state change. In contrast, the built-in pedal control did not have this feature, making it more susceptible to unintentional switching between states. Because the two interfaces differed in both input modality and controller logic (hysteresis), the observed difference cannot be attributed to trunk control alone. This pattern suggests that mode separation through hysteresis may provide more reliable navigation of hybrid continuous-discrete control spaces \cite{jacques2013hybrid}.

For the lower error counts observed in the BoMI group on the first transfer trial, a major difference in the transfer task from the training task was that it had curved-path segments that required participants to guide the paper while continuously modulating sewing speed. This finding may reflect differences in initial adaptation to the concurrent demands of machine speed control and bimanual paper manipulation. One possible explanation is that the coordination of the trunk and the hands in the BoMI might have been more intuitive than the hand-foot coordination required in the pedal group. The BoMI required the coordination of the upper body, where kinematic redundancy may allow the trunk to move independently, but also synergistically with the hands. In contrast, hand-foot coordination may impose different coordination demands \cite{nakagawa2016difference}. Therefore,  the difference observed on the first transfer trial may be specific to the coordination demands of this task. However, these coordination strategies were not measured directly and the difference between the groups were smaller over subsequent transfer trials. Furthermore, the transfer test  did not have established psychometric reliability measures and we did not have a baseline measure of transfer performance. Therefore, these findings should be considered secondary and require further validation in future studies.

\subsection{Implications}
These findings have implications for both the BoMI literature and accessible machine design. For the BoMI field, this study demonstrates that BoMIs can function effectively not only as standalone control systems, but also as one control channel within a coordinated multi-modal task. These results highlight that users  may be able to 'partition' their movement repertoire across the different subtasks and learn to execute these motions concurrently without any detectable task performance cost or elevated subjective workload.
Recent work has identified partitioning of degrees of freedom as a key strategy for controlling high-DOF assistive devices~\cite{ranganathan2026control}; 
the present findings suggest that such partitioning may also occur between device control and a concurrent manual task, not just across multiple device DOFs. This ability to use alternative control during concurrent manual activity highlights the potential for using BoMIs in real-world occupational applications.

In terms of accessible design, the results demonstrate that a single-IMU trunk-flexion interface can replicate the full functional repertoire of the foot with similar performance during a brief training session with unimpaired novices. This contribution is meaningful because foot pedal-operated sewing controls can be a direct access barrier for users with lower-limb impairments. In spite of  some progress in terms of developing alternative interfaces \cite{curry1993, mani2019development}, sewing remains highly ability-demanding on the user because fully automated solutions have to be extremely sophisticated to handle variations in cloth type, stitch complexity, and cloth quality. The BoMI approach allows the user to control the sewing machine through a lightweight wearable sensing setup, without requiring mechanical redesign of the sewing machine itself. Moreover, the flexibility of the BoMI to be tailored to a specific individual's movement repertoire \cite{farshchiansadegh2014body}, and adapt online to the user \cite{de2020guiding} likely paves the way for making sewing  accessible to a wide range of motor impairments.

\subsection{Limitations}
There are a few limitations associated with this study. First, we tested only healthy young adults without motor impairments. Although proof-of-concept studies are useful, testing with the target user population is an important consideration, especially given the focus on concurrent execution of multiple motor actions in individuals with limited movement repertoire  \cite{ranganathan2026control, biddiss2007consumer, beaudoin2019long}. Other factors such as postural control, fatigue, and proprioceptive acuity may also play a role in the success of the BoMI for users with motor impairments. Second, given the focus on learning, our study tested novices on this task over a single session. With a limited number of training trials, learning-rate and asymptote parameters were not estimated explicitly. Although other work has shown that BoMI control is retained well over longer durations\cite{pierella2015remapping, lee2024}, the question of whether the BoMI interface has the same learning curve as the pedal interface over longer time scales is still unknown.  Relatedly, the sample size of 12 participants per group, while consistent with prior BoMI studies \cite{lee2024,rizzoglio2023},  had limited statistical power to  detect small differences (or establish equivalence), and warrants replication in larger cohorts.

{Other limitations are related to the protocol and measures. Sewing on paper without thread in the Blum's test allowed for controlled, objective error scoring, but also meant that we did not assess other aspects of sewing such as seam integrity, cloth transport, material deformation, thread handling, or production-rate demands. 
In addition, repeated trunk movement may impose musculoskeletal demands not captured by NASA-TLX; direct measures of biomechanical loading, discomfort, and fatigue are needed to assess this. Future evaluation should involve threaded fabric, longer sessions, delayed retention and transfer testing, and ergonomic and safety endpoints, in participants who are unable to reliably use a pedal interface.
\par\medskip}

\subsection{Conclusion}
Overall, this study adds to the evidence that a BoMI interface can yield similar  performance relative to a conventional pedal control interface in a task requiring simultaneous machine control and bimanual manual work. Additional analyses suggest that BoMI control may offer task-specific benefits, as reflected in fewer presser-foot lift errors during training and fewer errors on the first transfer trial. These findings show that BoMI-mediated control can be effective not only in isolated device-control paradigms but also in multi-demand tasks. This represents an important step toward expanding the use of body-machine interfaces beyond traditional assistive devices contexts and into workplace settings.

\section*{Data Availability}
The data from this study, together with the processing and analysis scripts, are available at

\href{https://github.com/ashrut-2000/Sewing-BoMI}{https://github.com/ashrut-2000/Sewing-BoMI}

\bibliographystyle{IEEEtran}
\bibliography{references}

\end{document}